\documentclass[twocolumn,aps,prx,10pt,amsmath,amssymb,floatfix]{revtex4-1}

\usepackage{graphicx}
\usepackage{braket}
\usepackage{hyperref}
\usepackage{txfonts}
\hypersetup{colorlinks=true, linkcolor=blue, citecolor=blue, urlcolor=blue}

\def\doublestar{{}^\ast_\ast}
\def\doubledot{{}^{\cdot}_\cdot}
\newcommand{\no}[1]{\doubledot #1 \doubledot }
\newcommand{\nof}[1]{\doublestar #1 \doublestar}
\newcommand{\hc}{\text{h.c.}}

\begin{document}

\title{Bosonization for fermions and parafermions}

\date{\today}

\author{Thomas L.~Schmidt}
\affiliation{Physics and Materials Science Research Unit, University of Luxembourg, L-1511 Luxembourg}
\email{thomas.schmidt@uni.lu}

\begin{abstract}
Parafermions are fractional excitations which can be regarded as generalizations of Majorana bound states, but in contrast to the latter they require electron-electron interactions. Compared to Majorana bound states, they offer richer non-Abelian braiding statistics, and have thus been proposed as building blocks for topologically protected universal quantum computation. In this review, we provide a pedagogical introduction to the field of parafermion bound states in one-dimensional systems. We present the necessary theoretical tools for their study, in particular bosonization and the renormalization-group technique, and show how those can be applied to study parafermions.
\end{abstract}

\maketitle

\section{Introduction}

In the context of topological phases, Majorana bound states have been one of the most exciting research fields of the past decade \cite{alicea12,beenakker13,leijnse12}. They can be experimentally realized, for instance, by coupling either the helical edge state of a two-dimensional topological insulator \cite{deacon17,bocquillon16}, or a nanowire with Rashba spin-orbit coupling \cite{mourik12,albrecht16,gul18}, to a superconductor via the proximity effect. Aside from the fundamental interest in a new particle species, this research derives much of its relevance from the enormous potential of Majorana bound states for topological quantum computation \cite{dassarma06,nayak08,lahtinen17}. Indeed, qubits can be encoded in the degenerate ground states furnished by a collection of Majorana bound states, and thanks to their non-Abelian exchange statistics, certain qubit operations can be performed in a topologically protected way by braiding them.

However, Majorana bound states do not exhaust the full potential of topologically nontrivial bound states, and in recent years, important results have been obtained about their fractionalized cousins, the parafermionic bound states \cite{alicea16}. A set of operators $\chi_1, \ldots, \chi_N$ is said to satisfy $\mathbb{Z}_n$ parafermionic exchange statistics if
\begin{align}\label{eq:parafermion_def}
    \chi_j \chi_k = e^{2\pi i/n} \chi_k \chi_j, \quad \chi_j^n = 1, \quad \chi_j^\dag = \chi_j^{n-1}
\end{align}
for $1 \leq j < k \leq N$. According to this definition, Majorana bound states are $\mathbb{Z}_2$ parafermions. However, whereas Majorana bound states can exist in noninteracting systems, $\mathbb{Z}_n$ parafermions with $n \geq 3$ do require electron-electron interactions. This is why most proposals for their realization are based on strongly correlated phases such as fractional quantum Hall systems as host materials \cite{lindner12,clarke13,barkeshli13,mong14,barkeshli14,chen16,ebisu17,alavirad17}, but nanowires \cite{klinovaja13,klinovaja13b,sagi14,kainaris15,schmidt16,pedder16,pedder17,kornich17} and helical edge states \cite{sela11,klinovaja14,zhang14,orth15,hevroni16,traversoziani17,vinkleraviv17,fleckenstein19,strunz19} with interactions have been proposed as well.

The most important features of parafermions and some of their proposed experimental realizations have already been discussed in several excellent recent reviews \cite{alicea16,lahtinen17,nayak08}. Therefore, rather than giving an exhaustive overview of the recent literature, the aim of this text is to provide a pedagogical introduction to some of the theoretical tools necessary for the study of parafermions. Historically, parafermions originate from conformal field theory \cite{blumenhagen_book,senechal_book} but recent years have seen proposals for parafermionic bound states which are amenable to simpler theoretical methods. Since these proposals rely mostly on the technique of bosonization \cite{giamarchi03,senechal99,delft98}, this review will present the essential steps for using bosonization to study parafermions.

The bosonization identity is a remarkable exact mapping between fermionic and bosonic operators in one dimension \cite{delft98}. It is closely related to the Jordan-Wigner transformation and is at the heart of Luttinger theory \cite{luttinger63,tomonaga50}, which allows a description of one-dimensional gapless quantum systems at low energies for arbitrary interaction strength \cite{haldane81}. In fact, metallic one-dimensional systems have a large degree of universality and bosonization directly connects bosonic, fermionic and spin chains \cite{imambekov12}. The properties of such gapless one-dimensional systems have been the subject of many research projects over the past decades.

The aim of Luttinger theory is the description of \emph{gapless} systems. However, parafermionic bound states in 1D systems are typically bound to domain walls between two regions with topologically distinct spectral gaps. Hence, their study makes it necessary go beyond metallic 1D systems and embark on a deeper study of gapped systems. We will present the renormalization-group analysis as an instrument to find the conditions for opening gaps in 1D systems, and we will discuss different methods to study them. We will then proceed with a simple example on how to derive the parafermionic bound states from the bosonized expressions.

This review is structured as follows: in Sec.~\ref{sec:overview} we will present a brief, not necessarily exhaustive, summary of the recent research in this field. In Sec.~\ref{sec:bosonization}, we will discuss bosonization and in particular how to apply it for gapped one-dimensional systems. We will then move on to present a simple calculation for deriving the existence of parafermionic states in an example system in Sec.~\ref{sec:parafermions}. We will conclude the review in Sec.~\ref{sec:conclusion}.

\section{Overview}\label{sec:overview}

The simplest model containing Majorana bound states is the Kitaev chain \cite{kitaev2001}. It consists of spinless fermions hopping on a 1D lattice and coupled by $p$-wave superconducting pairing. Such a chain can be in two topologically distinct phases, and domain walls between different regions bind Majorana bound states. This prototypical model is related via a (nonlocal) Jordan-Wigner transformation to the more mundane quantum Ising model. The latter contains two competing terms, an exchange term $\propto S^z_{j} S^z_{j+1}$ and a perpendicular magnetic field term $\propto S^x_j$. The two topological phases of the Kitaev chain can be traced back to the ferromagnetic and paramagnetic phases of the quantum Ising chain, which are separated by a critical point at which the spectrum of the system becomes gapless.

An Ising spin on a given lattice site can point into one of two possible directions, corresponding to the states $\ket{\uparrow}$ and $\ket{\downarrow}$. To generalize this concept, one can define more general ``clock variables'', which owe their name to the analogy to a hand of a clock which is allowed to point in $n$ possible directions. The resulting model is called a $\mathbb{Z}_n$ clock model and can display similar critical points as the Ising model. The corresponding critical theories were first studied in 1985 \cite{zamolodchikov85} using conformal field theory, and parafermions were identified as their excitations. However, in those models, parafermions are not the only excitations and their connection to the lattice parafermions of Eq.~(\ref{eq:parafermion_def}) is nontrivial \cite{mong14a,alicea16}.

Read and Rezayi \cite{read99} demonstrated that quasihole states in quantum Hall systems at certain fractional filling factors (in particular $\nu = 12/5$ and $\nu=13/5$) obey parafermionic exchange statistics. This prediction was made possible by the preceding important insight that wave functions of fractional quantum Hall states are in fact closely related to correlation functions of certain conformal field theories \cite{moore91}. Nevertheless, these fractional filling factors are not very robust in experiments, and their excitations have not been studied experimentally so far.

One option to obtain interface bound states with parafermionic exchange statistics is to replace the noninteracting electrons which give rise to Majorana bound states in the Kitaev chain by fractionalized Laughlin quasiparticles. Laughlin states occur at fractional filling factors $\nu = 1/n$ with odd $n$, and are easier to observe experimentally. Starting out from a similar idea as in the Kitaev chain, namely from a 1D Laughlin edge state consisting of different regions which are gapped out by either superconductivity or some form of backscattering, one can show that the resulting interface bound states are $\mathbb{Z}_{2n}$ parafermions \cite{lindner12}. Several experiments have indeed shown in recent years that it is possible to induce superconductivity in quantum Hall edge states \cite{rickhaus12,amet16,lee17}.

In principle, such quantum Hall edge states can form the basis of more advanced parafermion chains. The resulting coupling of parafermionic interface states then leads to parafermionic lattices, which have been studied in several recent works \cite{fendley12,cobanera14,hutter15,hutter16,zhuang15,iemini17}. Moreover, it was shown that purely fermionic, Hubbard-like chain models can also give rise to exact parafermionic bound states, albeit with a less rich non-Abelian braiding statistics \cite{calzona18,mazza18,chew18,calvanesestrinati19}

In the limit of large system length, parafermions span a ground state space with a topologically protected degeneracy. This makes them ideal candidates for topological quantum computation \cite{lahtinen17}. Unfortunately, braiding is difficult, but not impossible, in a 1D system \cite{alicea11,lindner12}. Therefore, several theoretical proposals have studied the possibility of using fractional quantum Hall edge states as building blocks for more complicated 2D lattice structures. It has been shown that coupling between parafermions can then fuse them into Fibonacci anyons, whose braiding properties could allow universal quantum computation \cite{mong14}.

The fact that fractional quantum Hall edge states can host topologically protected parafermions is owed to a large extent to the fact that they are not true 1D systems, in the sense that such a chiral motion of quasiparticles is only possible because it happens at the edge state of a two-dimensional system. In contrast, several important theoretical works \cite{fidkowski11,turner11,chen11} have argued that even interacting strictly 1D systems should not carry more complex excitations than Majorana bound states. This, however, does not seem to entirely rule out interface states in 1D systems satisfying parafermionic exchange statistics. However, those interface states are then necessarily not fully topological, in the sense that a part of the ground state degeneracy can be lifted by local perturbations.

\section{Bosonization}\label{sec:bosonization}

Bosonization was derived as an exact mapping between fermionic and bosonic fields in one dimension \cite{mattis65}. The bosonization identity shows that chiral fermionic operators can be expressed as exponentials of bosonic operators. The usefulness of bosonization for more exotic states like para\-fer\-mions is a consequence of the fact that functions involving exponentials of bosonic operators can be used to construct quasiparticle operators with rather general exchange statistics, parafermions being one of them.

In the following, we will present a pedagogical example of how to construct parafermions in a simple 1D system. Our starting point will be the edge state of a two-dimensional topological insulator, where electrons with opposite spins travel in opposite directions. It has been known since the seminal work by Fu and Kane \cite{fu09b} that the superconducting proximity effect and a magnetic field generate spectral gaps with different topological invariants in this system, and that the interfaces between regions with different gaps host Majorana bound states. In the absence of electron-electron interactions, the edge state of a two-dimensional topological insulator can be described by the Hamiltonian,
\begin{align}\label{eq:Hel}
    H_{\rm el} = -i  v_F \int dx \left[ \nof{ \psi_R^\dag(x) \partial_x \psi_R(x) }  - \nof{ \psi_L^\dag(x) \partial_x \psi_L(x) }\right],
\end{align}
where $\psi_{R,L}(x)$ are field operators for right-moving spin-up (left-moving spin-down) electrons, and $v_F$ is the Fermi velocity. This Hamiltonian has a few noteworthy features: firstly, since right-movers and left-movers have opposite spins, the Nielsen-Ninomiya (``fermion doubling'') theorem implies that this Hamiltonian cannot emerge as the continuum limit of a 1D lattice Hamiltonian \cite{nielsen81}. Nevertheless it can describe edge electrons in topological insulator edge states because the host materials in this case are two-dimensional. Secondly, the operator $H_{\rm el}$ has time-reversal symmetry if the latter is defined as acting on the operators as $\Theta \psi_{R,L}(x) \Theta^{-1} = \pm \psi_{L,R}(x)$. Thirdly, we have linearized the spectrum and this has led to an infinite ``Dirac sea'', i.e., the spectrum of $H_{\rm el}$ is unbounded from below. To cure a trivial divergence of the ground state particle number, we have therefore used \emph{fermionic normal-ordering}, which is defined as,
\begin{align}
    \nof{ \psi_\alpha^\dag(x) \partial_x \psi_\alpha(x)} =  \psi_\alpha^\dag(x) \partial_x \psi_\alpha(x) - \langle \psi_\alpha^\dag(x) \partial_x \psi_\alpha(x) \rangle_0,
\end{align}
where $\alpha \in \{R, L\}$ and the subtracted quantity is the (divergent) expectation value with respect to the Dirac sea. However, it will turn out that the infinite spectrum of $H_{\rm el}$ has more subtle effects which will make it necessary to introduce a short-distance cutoff and bosonic normal ordering later.

Bosonization rests on identifying the fermionic fields $\psi_{R,L}(x)$ with bosonic fields. To set the stage for this transformation, we start with a seemingly unrelated Hamiltonian, namely that of a harmonic string, which is a sum of kinetic energy density and potential energy density,
\begin{align}\label{eq:Hbos}
    H_{\rm bos} = \frac{ v_s}{2\pi} \int dx \left\{ \Pi(x)^2 + [\partial_x \phi(x)]^2 \right\},
\end{align}
where $v_s$ is the speed of sound, and the momentum density and position operators are canonically conjugate, $[\partial_x \phi(x), \Pi(y)] = i \delta(x-y)$. This Hamiltonian can easily be diagonalized in terms of bosonic normal modes $b_p$, where $p = 2\pi n/L$ ($n \in \mathbb{Z}$) is the wave vector and $L$ is the length of the system. Note that $[b_p, b^\dag_{p'}] = \delta_{pp'}$. Defining a new bosonic operator $\theta(x)$ via $\partial_x \theta(x) = \pi \Pi(x)$, one finds that $H_{\rm bos} = \sum_p v_s |p| b_p^\dag b_p$ where \cite{giamarchi03},
\begin{align}
 \phi(x) &= -\frac{i \pi}{L} \sum_p \sqrt{\frac{L |p|}{2 \pi}} \frac{1}{p} e^{-a |p|/2} e^{-i p x} \left( b^\dag_p + b_{-p} \right), \\
 \theta(x) &= \frac{i \pi}{L} \sum_p \sqrt{\frac{L |p|}{2 \pi}} \frac{1}{|p|} e^{-a |p|/2} e^{-i p x} \left( b^\dag_p - b_{-p} \right).
\end{align}
To ensure convergence in the summation over momenta, it was necessary to introduce a short-distance cutoff $a$. The fields obey the commutation relation,
\begin{align}\label{eq:phi_theta_comm}
    [\phi(x), \theta(y)] = - \frac{i \pi}{2} \text{sgn}(x-y).
\end{align}
From these fields, it is straightforward to construct chiral right- and left-moving fields by defining $\varphi_{\alpha} = \alpha \phi - \theta$, where $\alpha = R, L = +, -$. The latter are indeed chiral because the Heisenberg equation of motion $\partial_t \varphi_\alpha(x,t) = i [H_{\rm bos}, \varphi_\alpha(x,t)]$ implies that these fields propagate only in one direction, $\varphi_\alpha(x,t) = \varphi_\alpha(x - \alpha v_s t)$.

At this point, we can introduce the bosonization identity, which provides an exact mapping between the chiral fermions fields $\psi_{R,L}$ and the chiral bosonic fields $\varphi_{R,L}$. The mapping is given by
\begin{align}\label{eq:bos_ident}
    \psi_{\alpha}(x) = \frac{1}{\sqrt{2 \pi a}} e^{-i \varphi_\alpha(x)} = \frac{1}{\sqrt{2 \pi a}} e^{-i \alpha \phi(x) + i \theta(x)},
\end{align}
where $a$ is again the same small distance cutoff as introduced before. A proof that bosonization is really an operator identity can be found in Ref.~\cite{delft98}. This identity also allows us to deduce that the time-reversal operator acts on the bosonic fields as $\Theta\phi(x) \Theta^{-1} = \phi(x) + \pi/2$ and $\Theta \theta(x) \Theta^{-1} = - \theta(x) + \pi/2$.

As a proof of principle, this expression already allows a calculation of the fermionic anticommutator. Splitting the chiral fields $\varphi_\alpha = \varphi_\alpha^+ + \varphi_\alpha^-$ into parts $\varphi^\pm_\alpha(x)$ containing only creation operators $b_p^\dag$ or only annihilation operators $b_p$, respectively, one finds the commutator,
\begin{align}\label{eq:varphi_comm}
   [\varphi^+_{\alpha}(x), \varphi^-_{\alpha'}(y)]
=
    \delta_{\alpha\alpha'} \ln \left( \frac{2\pi [a + i \alpha(x-y)]}{L} \right).
\end{align}
Using the Baker-Campbell-Hausdorff formula $e^{A} e^{B} = e^{A + B}e^{[A,B]/2}$ (valid in this form if $[A,B] \in \mathbb{C}$), and assuming that $L \gg x-y, a$, one can show that
\begin{align}
    \{ \psi_{\alpha}(x), \psi^\dag_\alpha(y) \} = \frac{1}{\pi} \frac{a}{a^2 + (x-y)^2} \stackrel{a \to 0}{=} \delta(x-y).
\end{align}
Two remarks about this result are in order. Firstly, in Eq.~(\ref{eq:bos_ident}), we neglected so-called Klein factors. The latter would in fact be necessary to find the vanishing anti-commutators $\{\psi_\alpha(x), \psi_{\beta}(x)\} = 0$ and $\{ \psi_L(x), \psi^\dag_R(y)\} = 0$. For many practical calculations, Klein factors are a trivial modification of the bosonization identity because they do not evolve in time and drop out when bosonizing expressions where the numbers of right- and left-movers is separately conserved, such as $\psi_\alpha(x) \psi^\dag_\alpha(y)$. We will ignore them in the following, but a more careful treatment of Klein factors is not difficult and can be found in Ref.~\cite{delft98}.

Secondly, one sees that the correct anticommutator is only obtained when taking the limit $a \to 0$ at finite $x-y$. It will remain true, even for bosonizing more complicated functions of fermionic fields, that $a$ must be assumed to be much smaller than the distance between the points $x$ and $y$ at which the field operators are evaluated. This means that effects at length scales smaller than $a$ cannot be described using bosonization, so the presence of a short-distance cutoff makes it an effective low-energy theory.

We have argued in this section that the bosonization identity Eq.~(\ref{eq:bos_ident}) can be used to translate between fermionic and bosonic operators. In the following sections, we will show why this is useful.

\subsection{Bosonic normal ordering}

When discussing the fermionic anticommutator, we have already encountered one of the problems which can arise when naively using the bosonization identity on products of fermionic operators evaluated at the same spacetime points, e.g., the density operator $\psi^\dag_R(x) \psi_R(x)$. Similar products occur in the kinetic energy operator, $\psi^\dag_R \partial_x \psi_R$. Bosonic normal ordering is a convenient technique to rigorously bosonize such expressions.

The bosonic operator $H_{\rm bos}$ has a unique vacuum state satisfying $b_p \ket{0} = 0$ for all $p$. This makes it possible to define bosonic normal ordering by the prescription of shifting all annihilation operators to the right, and all creation operators to the left. For example,
\begin{align}
    \no{b^\dag_{p_1} b_{p_2} b_{p_3} b^\dag_{p_4}} = b^\dag_{p_1} b^\dag_{p_4} b_{p_2} b_{p_3}.
\end{align}
Note that order of operators \emph{within} the product of creation operators (or annihilation operators) on the right hand side is irrelevant since they mutually commute. By construction, the vacuum expectation value of such a normal-ordered product vanishes.

Normal ordering is particularly important for exponentials of operators. For the operator occurring in the bosonization identity one finds using the binomial formula,
\begin{align}\label{eq:no_exp_tvarphi}
    \no{e^{i \lambda \varphi_\alpha}}
&=
    \sum_{n=0}^\infty \frac{(i\lambda)^n}{n!} \no{ (\varphi^+_\alpha + \varphi^-_\alpha)^n} \notag \\
&=
    \sum_{n=0}^\infty \sum_{m = 0}^n \frac{(i\lambda)^m (i\lambda)^{n-m} (\varphi^+_\alpha)^m (\varphi^-_\alpha)^{n-m}}{m! (n-m)!} \notag \\
&=
    e^{i \lambda \varphi^+_\alpha} e^{i \lambda \varphi^-_\alpha}.
\end{align}
This result also follows from the fact that linear functions of creation and annihilation operators, such as $\varphi^\pm_\alpha(x)$, commute under normal ordering. Next, we use again the Baker-Hausdorff formula to combine the exponents. Finally, taking the limit $L \gg x-y, a$, one finds~\cite{delft98}
\begin{align}\label{eq:no_exp}
    \no{e^{i \lambda \varphi_\alpha}}
=
     \left( \frac{L}{2 \pi a} \right)^{\lambda^2/2} e^{i \lambda \varphi_\alpha}.
\end{align}
Using the expansion $e^{i \lambda \varphi_\alpha} = 1 + i \lambda \varphi_\alpha + (\ldots)$, one can easily see that the vacuum expectation value of a normal-ordered exponential of a bosonic operator is always one: $\braket{\no{e^{i \lambda \varphi_\alpha(x)}}} = 1$. Therefore, we can also write $\no{e^{i \lambda \varphi_\alpha}} = e^{i \lambda \varphi_\alpha}/\braket{e^{i \lambda \varphi_\alpha}}$.
One can analogously derive an equation useful for normal ordering products of exponentials,
\begin{align}\label{eq:no_exp_varphi_pair}
&
    e^{i \lambda \varphi_\alpha(x)} e^{i \lambda' \varphi_\alpha(y)} \notag \\
&=
    \left( \frac{L}{2 \pi a} \right)^{-(\lambda^2+\lambda'^2)/2} \left( \frac{2 \pi [ a + i \alpha (y - x)]}{L} \right)^{\lambda \lambda'} \no{e^{i \lambda \varphi_\alpha(x) + i \lambda' \varphi_\alpha(y)}}
\end{align}
The advantage of normal ordering is that all cutoff-dependent and potentially divergent terms have been extracted from the operators. The limit $x \to y$ needs to be taken with care in the prefactor, but in the normal-ordered operator, it can be taken straightforwardly.

\subsection{Density and kinetic energy}

Using Eq.~(\ref{eq:no_exp_varphi_pair}), it becomes simple to directly derive bosonized expressions for bilinear fermionic operators. As a first example, we consider the fermionic density operators. We start from the product,
\begin{align}\label{eq:den_psipsi}
    \psi^\dag_\alpha(x) \psi_\alpha(y)
&=
    \frac{1}{2\pi a} e^{i \varphi_\alpha(x)} e^{-i \varphi_\alpha(y)}
=
    \frac{\no{e^{i \varphi_\alpha(x)}
        e^{- i \varphi_{\alpha}(y)}}}{2 \pi [a + i \alpha(y - x)]}
\end{align}
As before, we should take the limit $a \to 0$ first and afterwards perform a Taylor expansion to first order in $x - y$. Finally, one performs fermionic normal-ordering to eliminate a constant term. Then, we find for the normal-ordered fermionic density,
\begin{align}\label{eq:rho_boson}
    \nof{\psi^\dag_\alpha(x) \psi_\alpha(x)}
=
    - \frac{\alpha}{2\pi}
    \partial_x \varphi_\alpha(x).
\end{align}
After this warm-up exercise, let us investigate the fermionic kinetic energy in a similar fashion. We start by differentiating Eq.~(\ref{eq:den_psipsi}) with respect to $y$ and assume as before that $a \ll x - y$,
\begin{align}\label{eq:psidpsi_boson}
    \psi^\dag_\alpha(x) \partial_{y} \psi_\alpha(y)
&=
    \frac{1}{2 \pi} \bigg\{ \frac{i \alpha}{(y-x)^2}  \no{e^{i \varphi_\alpha(x)}
        e^{- i \varphi_{\alpha}(y)}} \notag \\
    &+  \frac{1}{i \alpha (y-x)}      \partial_{y} \no{e^{i \varphi_\alpha(x)}
        e^{- i \varphi_{\alpha}(y)}}  \bigg\}.
\end{align}
Next, we have to expand the exponentials up to the second and first order in $x-y$, respectively,
\begin{align}
    \no{e^{i \varphi_\alpha(x)}
        e^{- i \varphi_{\alpha}(y)}}
&\approx
    \no{1 - i (y-x) \varphi_\alpha'(x) \notag \\
&-\frac{i  (y-x)^2}{2} \varphi_\alpha''(x)- \frac{(y-x)^2}{2} (\varphi_\alpha')^2}, \notag \\
    \no{e^{i \varphi_\alpha(x)} \partial_{y} e^{- i \varphi_{\alpha}(y)}}
&\approx
    \no{-i \varphi_\alpha'(x) - (y-x) [\varphi_\alpha'(x)]^2}.
\end{align}
Note that the constant term in the first line drops out when using fermionic normal ordering. We now consider the kinetic Hamiltonian $H_{\rm el}$ in Eq.~(\ref{eq:Hel}) for a linear spectrum. We first rewrite it in a hermitian form,
\begin{align}
    H_{\rm el}
&= \sum_\alpha \frac{\alpha  v_F}{2}  \int dx \left[ -i\ \nof{\psi_\alpha^\dag(x) \partial_x \psi_\alpha(x) }   + \hc \right].
\end{align}
Taking into account that the fields $\varphi_\alpha(x)$ are hermitian, we find the well-known result
\begin{align}\label{eq:Hel_bos}
    H_{\rm el} &= \frac{ v_F}{4\pi} \sum_\alpha  \int dx \no{[\partial_x \varphi_\alpha(x)]^2} \notag \\
&=
    \frac{ v_F}{2\pi} \int dx \left\{ \no{[\partial_x \theta(x)]^2} + \no{[\partial_x \phi(x)]^2} \right\},
\end{align}
showing that the noninteracting fermionic Hamiltonian (\ref{eq:Hel}) is identical to the bosonic Hamiltonian (\ref{eq:Hbos}) if we identify the Fermi velocity $v_F$ with the sound velocity $v_s$.

Using bosonization it becomes almost trivial to incorporate the effect of electron-electron interactions. The essential, and perhaps surprising, insight is that, as shown in Eq.~(\ref{eq:rho_boson}), the bilinear fermionic density operator is proportional to a linear bosonic operator. This means that density-density interactions, being quartic in fermionic operators, become quadratic in terms of bosonic operators.

If the electron-electron interactions are weak, one can use Eq.~(\ref{eq:Hel}) as a starting point and project the fermionic interaction Hamiltonian onto the basis of right-movers and left-movers near the Fermi points. Then, one mainly needs to consider two types of interactions,
\begin{align}\label{eq:int1}
    ( \psi^\dag_\alpha \psi_\alpha ) ( \psi^\dag_\alpha \psi_\alpha ) &\propto \no{(\partial_x \varphi_\alpha)^2} \notag \\
    ( \psi^\dag_L \psi_L ) ( \psi^\dag_R \psi_R ) &\propto \no{(\partial_x \varphi_L) (\partial_x \varphi_R)}
\end{align}
The first line involves scattering processes with momentum exchange $|q| \ll k_F$, whereas the processes in the second line can happen either with $|q| \approx 2k_F$ or with $|q| \approx 0$. Combining these interaction terms with the bosonized version of the kinetic Hamiltonian (\ref{eq:Hel_bos}), one finds a result of the form,
\begin{align}\label{eq:Hbos2}
    H_{LL} = \frac{v}{2\pi} \int dx \left[ K \no{ (\partial_x \theta)^2} + \frac{1}{K} \no{(\partial_x \phi)^2} \right]
\end{align}
The ``Luttinger liquid'' Hamiltonian $H_{LL}$ differs from Eq.~(\ref{eq:Hel_bos}) in two important points: a change of the effective sound velocity $v$ and the emergence of the so-called Luttinger parameter $K$. For weakly interacting fermions, $K$ and $v$ are determined by the (unperturbed) Fermi velocity $v_F$ as well as by the Fourier components near $q \approx 0$ and $|q| \approx 2k_F$ of the interaction potential between the physical electrons. Fortunately, however, the validity of the Hamiltonian (\ref{eq:Hbos2}) is not limited to weakly interacting fermions: it was shown by Haldane \cite{haldane81} that Eq.~(\ref{eq:Hbos2}) can describe all gapless 1D quantum systems at low energies if one considers $K$ and $v$ as phenomenological parameters.

Interactions have a subtle effect on normal ordering as well. The operator $H_{LL}$ is bilinear, but not diagonal in the operators $b_p$ and $b_p^\dag$. Indeed, since it contains ``pairing'' terms proportional to $b_p b_{-p}$ and $b^\dag_p b^\dag_{-p}$, it has to be diagonalized using a Bogoliubov transformation, which leads to new eigenmodes $\tilde{b}_p = f(b_p, b^\dag_{-p})$. This means, however, that the respective vacuum states of $b_p$ and $\tilde{b}_p$ will differ for $K\neq 1$. Therefore, we will assume in the following that normal ordering is understood to be with respect to the eigenmodes $\tilde{b}_p$ of the interacting system.

It is worth pointing out that the Luttinger Hamiltonian (\ref{eq:Hbos2}) can also be used to model fractional quantum Hall edge states of the Laughlin sequence if one takes into account that for filling factor $\nu = 1/n$ (with odd $n$) the canonical commutation relation (\ref{eq:phi_theta_comm}) between the phase fields is modified to $[\phi(x), \theta(y)] = -i \pi \text{sgn}(x-y)/(2n)$.

\subsection{Renormalization group analysis}

\newcommand{\tg}{\tilde{g}}
\newcommand{\tphi}{\tilde{\phi}}
\newcommand{\tpsi}{\tilde{\psi}}
\newcommand{\trho}{\tilde{\rho}}
\newcommand{\ttheta}{\tilde{\theta}}
\newcommand{\tDelta}{\tilde{\Delta}}

A renormalization group (RG) analysis is one of the most important tools for analyzing correlated systems \cite{cardy96,shankar94}. By systematically integrating out high-energy degrees of freedom, it will allow us to derive effective low-energies Hamiltonians from which one can then determine the possible phases of Hamiltonians in the presence of different kinds of interaction terms. In view of parafermions in helical edge states (of a topological insulator or a quantum Hall system), the most important interaction terms which drive transitions to topological phases are superconducting pairing and backscattering of single particles or particle pairs.

Single-particle backscattering corresponds to an operator $\propto \psi^\dag_R \psi_L$. In the helical edge state of a topological insulator, right- and left-movers have opposite spins, so this term can be generated by a magnetic field applied perpendicularly to the spin quantization axis. However, single-particle backscattering changes the total momentum by $2k_F$, so it is kinematically allowed only if the chemical potential is near the Dirac point ($k_F = 0$). Bosonizing this term is straightforward and leads to
\begin{align}
    H_{1bs} = \frac{g_{1}}{2\pi a} \int dx \cos[ 2 \phi(x) ].
\end{align}
$H_{1bs}$ violates time-reversal symmetry because $\Theta H_{1bs} \Theta^{-1} = - H_{1bs}$. The combination of electron-electron interactions and spin-orbit coupling can also give rise to two-particle backscattering, described by interaction terms $H_{2bs} \propto \psi^\dag_R ( \partial_x \psi^\dag_R) (\partial_x \psi_L) \psi_L + \hc$. It is time-reversal symmetric but still breaks spin conservation, which is why spin-orbit coupling is essential \cite{orth15}. Note that the derivatives of the fields appear naturally due to the Pauli principle: $H_{2bs}$ can be regarded as the continuum limit of a lattice backscattering term which acts on fermions on neighboring lattice sites.

As $H_{2bs}$ involves derivatives, bosonizing it requires careful normal ordering. As in Eq.~(\ref{eq:psidpsi_boson}), one starts by bosonizing $H_{2bs}$ assuming that all four fermionic fields are located at different positions $x_{1,2,3,4}$. Afterwards, one takes the limit $x_{1,2,3,4} \to x$ in the completely normal ordered expression. The result then is
\begin{align}
    H_{2bs} = \frac{g_{2}}{2\pi a} \int dx \cos[ 4 \phi(x) ].
\end{align}

Let us assume we start with an electronic system with density-density interactions, to which we add one of the terms corresponding to backscattering. Therefore, we start from the general Hamiltonian,
\begin{align}
    H_{LL} &= \frac{v}{2 \pi} \int dx \left[ \no{(\partial_x \ttheta)^2} + \no{(\partial_x \tphi)^2} \right] \notag \\
    H_{g} &= \frac{g}{2 \pi a} \int dx \cos[2 \lambda \phi(x)] = \frac{v \tg}{2 \pi a^2} \int dx \cos[2 \lambda \sqrt{K} \tphi(x)] \notag \\
&=
    \frac{v \tg}{2 \pi a^2} \left(\frac{2\pi a}{L}\right)^{\lambda^2 K} \int dx \no{\cos[2 \lambda \sqrt{K} \tphi(x)]}
\end{align}
where we defined $\tphi = \phi/\sqrt{K}$ and $\ttheta = \sqrt{K} \theta$ to diagonalize $H_{LL}$. With this definition, $g$ has the dimension of energy and $\tg = a g/v$ is dimensionless. The model contains a short-distance cutoff $a$. The main idea of an RG analysis is to investigate the dependence of the coupling parameter $\tg$ on such a cutoff. Normal-ordering is convenient for such an RG analysis as it allows us to extract the cutoff dependence of an operator.

Perturbative RG rests on the principle that physical quantities should remain invariant under the choice of cutoff, provided that the coupling constants are changed accordingly. A possible starting point is the $S$ matrix in the interaction picture \cite{mahan90}
\begin{align}
    S = T \exp\left[ -i \int_{-\infty}^\infty dt V(t) \right],
\end{align}
where $V(t)$ is the perturbation operator with time evolution governed by the unperturbed Hamiltonian $V(t) = e^{i H_0 t} V e^{-i H_0 t}$. Moreover, $T$ denotes the time-ordering operator. In our case, we take $V = H_g$ and $H_0 = H_{LL}$. Since we are using perturbation theory, we can expand this for small $\tg$,
\begin{align}
    S
&\approx 1 + S_g^{(1)} + S_g^{(2)} + \ldots \notag \\
&= 1 - i \int dt H_g(t) - \frac{1}{2} \int dt_1 dt_2 T H_g(t_1) H_g(t_2).
\end{align}
To investigate the RG flow, we consider $S$ as a function of $a$ and $\tg$. Demanding its invariance under change of cutoff, our aim is to find a renormalization of the coupling constant in such a way that
\begin{align}
    S(a, \tg) = S(a - da, \tg + d\tg),
\end{align}
where $da < 0$. Since all operators are normal-ordered, we can directly read off the cutoff dependence. The first order term gives us the condition $S_g^{(1)}(a, \tg) = S_g^{(1)}(a - da, \tg + d\tg)$. Since both sides contain the same integral, we simply have to compare the prefactors,
\begin{align}
&
    - \frac{i v \tg}{2 \pi a^2} \left(\frac{2\pi a}{L}\right)^{\lambda^2 K} = - \frac{i v (\tg + d\tg)}{2 \pi (a - da)^2} \left(\frac{2\pi (a - da)}{L}\right)^{\lambda^2 K},
\end{align}
which leads to the scaling equation $d\tg = (\lambda^2 K-2) \tg da/a$. We parameterize the cutoff as $a(\ell) = a_0 e^{-\ell}$, where $a_0$ is some initial cutoff. This leads to $da = - a d\ell$ and to the flow equation
\begin{align}\label{eq:RG_1}
    \frac{d\tg}{d\ell} = (2-\lambda^2 K) \tg.
\end{align}
By construction, increasing the short-distance cutoff, i.e., to going to lower energy scales, corresponds to a positive $d\ell > 0$. Hence, the flow equation tells us that, depending on $\lambda^2 K$ the coupling constant $\tg$ may increase, decrease or stay invariant. In the former case, i.e., for $K < 2/\lambda^2$, the perturbation is called RG-relevant. If the RG flow can be continued the zero energy, the effective coupling constant will then increase exponentially, and the phase at low energies will be determined by the coupling Hamiltonian $H_g$.

If, on the other hand, one finds $d\tg/d\ell < 0$, the perturbation is said to be irrelevant and can be treated as a small perturbation at low energies. The case where $d\tg/d\ell = 0$ is called marginal. In this case, the relative importance of $H_{LL}$ and $H_g$ remains identical under the change of cutoff, so the RG analysis, at least up to this order, yields no answer regarding the low-energy phase, and a second-order RG analysis is necessary. Such a second-order treatment also leads to a renormalization of the Luttinger parameter $K$ and reveals that the phase transition between the phase where $H_g$ is RG-irrelevant and the phase where $H_g$ is RG-relevant is of Kosterlitz-Thouless type \cite{giamarchi03}.

\subsection{Spectral gaps}

What happens when a cosine term becomes RG-relevant? The short answer is that the system becomes gapped, but in fact there are different, equally interesting ways to see this. Adding the cosine term breaks the conformal symmetry of $H_{LL}$ but leaves it Lorentz-invariant. This restricts the spectrum to the Lorentzian form,
\begin{align}
    E(k) = \sqrt{(v k)^2 + E_g^2}
\end{align}
with some gap energy $E_g$. For a Luttinger liquid, $E_g = 0$, but in the presence of an RG-relevant cosine term, $E_g$ is proportional to its strength.

The ``sine-Gordon'' model $H_{SG} = H_{LL} + H_{g}$ is one of the few exactly solvable models of quantum field theory \cite{smirnov_book}. It was known for a long time that the classical equation of motion of the model is the sine-Gordon equation, an exactly solvable nonlinear differential equation whose solutions are solitons. It was shown much later that the $S$ matrix of the quantum field theoretical model can be found exactly \cite{zamolodchikov79,zamolodchikov95}, and even form factors (matrix elements of operators in the eigenstates) can be determined \cite{lukyanov97a,lukyanov01}. However, in the presence of competing cosines, for instance due to a simultaneous presence of a superconducting term and a backscattering term in different regions, the model is not integrable any more. Moreover, the solitons are excited states of the sine-Gordon model, i.e., with energies about the gap, and therefore in general not necessary to describe topological phases.

Alternatively, the phase diagram as well as some response functions of $H_{SG}$ can be obtained using refermionization. If the cosine term is the most RG-relevant term, it will dominate the low-energy phase, so it is judicious to choose a basis which diagonalizes this term. One starts by defining the rescaled bosonic fields $\phi' = \lambda \phi$, $\theta' = \theta/\lambda$, which remain canonically conjugate. As a consequence, the definition
\begin{align}\label{eq:referm}
    \psi'_{\alpha}(x) \propto e^{-i \alpha \phi' + i \theta'}
\end{align}
ensures that the fields $\psi'_\alpha(x)$ mutually anticommute. Reverting the steps leading to the bosonized Hamiltonian, one can then show that
\begin{align}
    H_{SG}
&= -i v K \lambda^2 \int dx \left[ \nof{\psi_R'^\dag \partial_x \psi'_R} - \nof{ \psi_L'^\dag \partial_x \psi'_L} \right] \notag \\
&+ \frac{g}{2} \int dx \left( \psi_R'^\dag \psi'_L + \hc \right) \notag \\
&+
    \frac{\pi v_s}{2} \left(\frac{1}{\lambda^2 K} - \lambda^2 K \right) \int dx \left[\rho'_L(x) + \rho'_R(x)\right]^2.
\end{align}
Here, $\rho'_\alpha(x) = \nof{\psi'^\dag_\alpha(x) \psi'_\alpha(x)}$ is the quasiparticle density. This model is called the massive Thirring model and its equivalence to the sine-Gordon model was in fact known long before bosonization \cite{coleman75}. In this form, the term proportional to $g$ makes it obvious that the spectrum is gapped, but apart from that it might seem that little has been gained: the refermionized model contains interactions between the quasiparticles. Luckily however, these interactions are ``weak''. In an RG-sense, they constitute a marginal correction to the kinetic energy Hamiltonian, whereas the $g$ term is still an RG-relevant correction. Therefore, perturbative techniques can be employed. One very prominent technique is a ladder-diagram resummation which is well-known in the contact of the Mahan exciton problem \cite{mahan90}.

It is worth pointing out as well that the refermionized model has an exactly solvable point where the quasiparticles become non-interacting ($K=1/\lambda^2$ in the example above). Such a point is called a Luther-Emery point \cite{luther74a}. This special point is useful to get an insight into the phase diagram, but it should be pointed out that response functions, especially close to gap energies, change strongly in response to quasiparticle interactions \cite{mahan90}. The exact solution of the sine-Gordon model indeed reveals that the Luther-Emery point is a special point and that deviations from this point lead to non-analytical corrections to response functions.

The most user-friendly approximation to study sine-Gordon Hamiltonians relies on the argument that an RG-relevant cosine term will ``pin'' the phase field to a minimum of the cosine function. For the Hamiltonian $H_g$ above this would mean that one can approximate $\tphi \approx \pi/(2 \lambda \sqrt{K})$. Performing a Taylor expansion $\tphi \approx \tphi_{\rm min} + \delta \tphi$ about one of the possible minima yields a quadratic Hamiltonian in $\delta \tphi$. The latter can be solved exactly. It gives direct access to the spectrum, and can in principle be used to calculate response functions. However, it is worth pointing out that this ``pinning'' approximation is not very accurate when it comes to dynamic correlation functions of sine-Gordon models, as can be shown by comparing with the exact solution based on the form factors of the sine-Gordon model.

Modelling sine-Gordon terms using this ``pinning'' approximation is an excellent approximation for studying ground-state properties, which allows us to use it below for deriving the edge parafermion modes. We should point out that another very fruitful application of sine-Gordon Hamiltonians is the wire constructions which are presented by T.~Meng in the same special issue \cite{meng19}.

\section{Parafermions}\label{sec:parafermions}
\newcommand{\tot}{{\rm tot}}
\newcommand{\hn}{\hat{n}}
\newcommand{\hm}{\hat{m}}

In this section, we will show how the competition between different non-commuting, RG-relevant sine-Gordon terms in a bosonized Hamiltonian can lead to the emergence of parafermionic bound states. In 1D systems, parafermions are bound to domain walls at the interfaces between parts of the system with topologically different spectral gaps. In the following, we will choose the specific example of a time-reversal invariant helical edge state of a 2D topological insulator, where different regions are gapped out either by the superconducting proximity effect or by two-particle backscattering. In this case, the resulting interface bound states are $\mathbb{Z}_4$ parafermions \cite{zhang14,orth15}. We would like to point out, however, that the procedure we will present is more general, and has been used for studying $\mathbb{Z}_{2n}$ parafermions in fractional quantum Hall systems with filling factor $1/n$ (with odd $n$) as well \cite{lindner12,clarke13}.

We have already presented the Hamiltonian $H_g$. Another potentially RG-relevant term which can be added to the helical edge Hamiltonian results from the proximity effect in the presence of a superconductor,
\begin{align}
    H_{\rm sc} &= \Delta \int dx \left[ \psi_R^\dag(x) \psi_L^\dag(x) + \hc \right]
=
    \frac{\Delta}{2\pi a} \int dx \cos[2 \theta(x)] \notag \\
&=
    \frac{v \tDelta}{2\pi a^2} \int dx \cos[2 \ttheta(x)/\sqrt{K}] \notag \\
&=
    \frac{v \tDelta}{2\pi a^2} \left( \frac{2\pi a}{L} \right)^{2/K} \int dx \no{\cos[2 \ttheta(x)/\sqrt{K}]}.
\end{align}
Here, $\Delta$ denotes the induced gap energy, which is a function of the bulk superconducting gap as well as of the electron tunneling amplitude between superconductor and the edge state, and $\tDelta$ is its dimensionless version. As before, we obtain the flow equation
\begin{align}
    \frac{d\tDelta}{d\ell} = \left( 2- \frac{1}{K} \right) \tDelta.
\end{align}
From this, we can conclude the superconductivity is relevant for $K > 1/2$. Clearly, the RG flow of cosine terms containing $\theta$ or $\phi$ change in opposite ways as a function of $K$: gaps due to backscattering are typically reinforced by strong repulsive interactions, whereas superconducting coupling is inhibited by them.

The basic ideas for deriving parafermionic operators are as follows: based on the assumption that each RG-relevant cosine term pins its bosonic field to a minimum, one can identify a finite set of ground states. Then, one constructs operators, localized at the interfaces, which cycle between those degenerate ground states. The latter are the desired parafermionic bound states.

\begin{figure}
\centering
\includegraphics[width=\columnwidth]{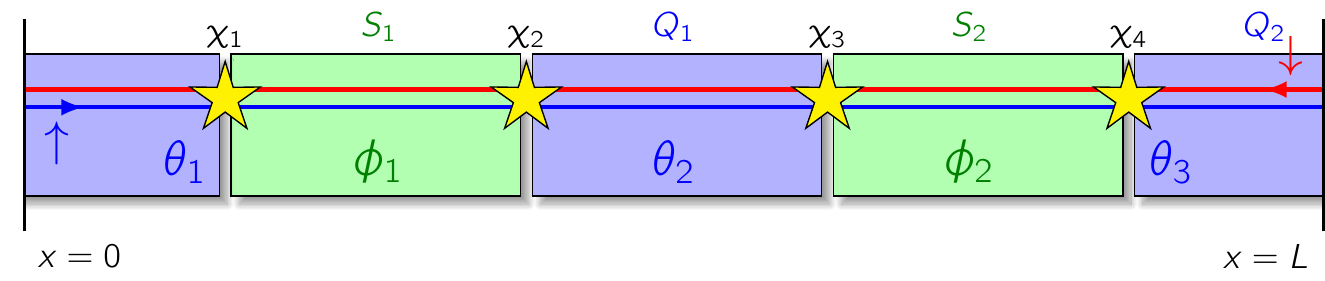}
\caption{Alternating superconducting (blue) and backscattering (green) regions with parafermionic interface states indicated by stars. In the superconducting (backscattering) region, the field $\theta$ ($\phi$) is pinned.}
\label{fig:chain}
\end{figure}

To be concrete, let us consider a minimal model consisting of a helical edge state, in which one region is gapped out by superconductivity, whereas another one is gapped out by umklapp scattering. This corresponds to the Hamiltonian $H = H_{LL} + H_{g} + H_{sc}$, where
\begin{align}
    H_{LL} &= \frac{v}{2\pi} \int dx \left[ K \no{(\partial_x \theta)^2} + \frac{1}{K} \no{(\partial_x \phi)^2} \right], \notag \\
    H_{g} &= -\int dx g(x) \cos[4 \phi(x) ], \notag \\
    H_{sc} &= -\int dx \Delta(x) \cos[2 \theta(x) ].
\end{align}
Near $K = 1/2$ both terms can open spectral gaps. In the respective regions, the cosine terms will pin the phase field $\theta$ and $\phi$ as shown in Fig.~\ref{fig:chain} for case of $N=2$ junctions. Without loss of generality, we can assume that $g > 0$ and $\Delta > 0$. In the superconducting regions, $\theta(x)$ will be pinned to values $\theta(x) = \hn \pi$, whereas backscattering will pin the $\phi$ field to values $\phi(x) \in \hm \pi/2$, where $\hn$ and $\hm$ are now operators with integer spectrum, who inherit their commutation relations from those of the phase fields [see Eq.~(\ref{eq:phi_theta_comm})].

Next, we define operators corresponding to the total charge and the total spin in a given region $x_1 < x < x_2$. From Eq.~(\ref{eq:rho_boson}), one obtains
\begin{align}
    Q(x_1,x_2)
&=
    \sum_{\alpha = R,L} \int_{x_1}^{x_2} dx \nof{ \psi^\dag_\alpha(x) \psi_\alpha(x) } \notag \\
&=
    -\frac{1}{\pi} \left[ \phi(x_2) - \phi(x_1) \right], \notag \\
    S(x_1,x_2)
&=
    \sum_{\alpha = R,L} \alpha \int_{x_1}^{x_2} dx \nof{ \psi^\dag_\alpha(x) \psi_\alpha(x) } \notag \\
&=
    -\frac{1}{\pi} \left[ \theta(x_2) - \theta(x_1) \right].
\end{align}
Due to the pinning of the fields by the cosine terms, one finds that the spin trapped between two superconducting regions, as well as the charge trapped between two regions gapped by backscattering, are quantized. Note that while the spin is quantized to integer multiplies of the electron spin, the charge is quantized to half-integer multiples of the elementary charge (see Fig.~\ref{fig:chain} for the notation),
\begin{align}
    S_n &= - (\theta_{n+1} - \theta_n)/\pi = - \hn_{n+1} + \hn_n, \notag \\
    Q_n &= - (\phi_{n+1} - \phi_n)/\pi = - (\hm_{n+1} - \hm_n)/2.
\end{align}
The half-integer quantization can most easily be understood by using the refermionization transformation (\ref{eq:referm}) on the backscattering term. Using $\theta' = \theta/2$ and $\phi' = 2\phi$, one finds that $\psi^\dag_\alpha \propto (\psi'^\dag_\alpha)^2 e^{-3 i \alpha \phi}$. Since $\phi$ commutes with the charge operator, it does not create charge. Therefore, charge is only created by the $\psi'^\dag_\alpha$ operator, which reveals that the latter carries half an electron charge.

To learn more about the ground state degeneracy, one first needs to determine the domain of the fields $\phi(x)$ and $\theta(x)$. Since each superconducting region conserves charge modulo $2$, whereas each backscattering region conserves spin modulo $4$, one finds that there are four possible values for each $S_n$ and $Q_n$, so $S_n \in \{0,1,2,3\}$ and $Q_n \in \{ 0, \tfrac{1}{2}, 1, \tfrac{3}{2} \}$. This argument can be made more mathematically rigorous using translation operators for the phase fields \cite{orth15}.

The different possible values of the pinned phase fields cause a degeneracy of the ground state. The latter can be determined by constructing a complete set of commuting observables, where it has to be taken into account that $Q_n$ and $S_n$ do not mutually commute as a consequence of the nonlocal commutation relations (\ref{eq:phi_theta_comm}). Using the largest possible set of mutually commuting operators, one finds that for $2N$ interfaces each ground state can be parameterized as $\ket{s_1, \ldots, s_{N-1}, s_\tot, q_\tot}$, where $s_j$ are the eigenvalues of $S_j$, and $s_\tot$ that of the total spin operator $S_\tot = \sum_{j} S_j = \hn_1 - \hn_{N+1}$. Similarly, $q_\tot$ is the eigenvalue of the total charge operator $Q_\tot = (\hm_1 - \hm_{N+1})/2$.

In contrast to the interface charges, the total charge of the entire system must be integer, so $q_\tot \in \{ 0,1 \}$. Analogously, the total spin must reflect the fact that each electron carries one unit of spin, so $s_\tot \in \{0,2\}$ for $q_\tot = 0$ and $s_\tot \in \{1,3\}$ for $q_\tot = 1$. This leads to a total ground state degeneracy of $4^{N-1} \times 2 \times 2 = 4^{N}$.

Finally, we can construct the parafermion bound states operators. They should be local in the bosonic fields, i.e., any given bound state operator should only involve the pinned phase fields in the two adjacent regions. Moreover, all bound state operators should act within the ground state vector space. The resulting $2N$ operators are
\begin{align}
    \chi_{2n-1} &= e^{i \phi_n - i \theta_n/2}, \notag \\
    \chi_{2n} &= e^{i \phi_n - i \theta_{n+1}/2}.
\end{align}
Using the Baker-Hausdorff formula as well as the commutator (\ref{eq:phi_theta_comm}) of the bosonic fields, one can indeed verify the parafermionic commutation relations (\ref{eq:parafermion_def}).

The operators $\chi_n$ obey the anyonic commutation relations of parafermionic operators, but for them to be non-Abelian particles, the unitary transformation corresponding to their adiabatic exchange must be a representation of the braid group. Exchanging the positions of two parafermions in real space, without bringing them together, is not possible in a 1D edge state. Therefore, alternative braiding schemes have been proposed which rely on repeatedly nucleating and fusing pairs of parafermions \cite{lindner12}. This can be regarded as a braiding operation in parameter space, and it has been shown to be topologically protected in the sense that weak deformations of the braiding path do not change the braiding phase. However, this robustness is significantly weaker than that afforded by real-space braiding of parafermions in 2D systems, because certain phases in the cosine terms could cause accidental degeneracies which must be avoided \cite{lindner12,chen16}.

\section{Conclusions}\label{sec:conclusion}

In this review, we have discussed the emergence of parafermions in one-dimensional quantum systems, with a particular focus on explaining the theoretical techniques necessary for their modeling. Bosonization has for many years been the method of choice for investigating one-dimensional systems. For many gapless systems, the representation of fermionic particles in terms of bosonic density waves leads to an exact mapping between an interacting fermionic theory and a noninteracting bosonic one, and thus allows one to study the full crossover between the weakly and strongly interacting limits.

To construct parafermions, one needs to supplement the well-known metallic Luttinger Hamiltonian with gap-opening terms, which usually leads to sine-Gordon type Hamitonians. We've explained how a renormalization-group analysis reveals the possible phases of such Hamiltonian. In the strong-coupling limit, the cosine terms in the Hamiltonian cause a ground state degeneracy, and parafermionic operators emerge as the local operators which cycle between the ground states.

\section*{Acknowledgements}
The author acknowledges support by the National Research Fund, Luxembourg under grants ATTRACT 7556175, INTER 11223315, PRIDE/15/10935404, and AFR/11224460.

\bibliography{refsclean}

\end{document}